\documentstyle[prb,aps,multicol,epsf]{revtex}
\renewcommand{\narrowtext}{\begin{multicols}{2}
\global\columnwidth20.5pc}
\renewcommand{\widetext}{\end{multicols} \global\columnwidth42.5pc}
\def\Lrule{\vspace*{-0.2in}\noindent\vrule width3.5in height.2pt
  depth.2pt \vrule depth0em height1em}
\def\Rrule{\vspace{-0.1in}\hfill\vrule depth1em height0pt \vrule
  width3.5in height.2pt depth.2pt\vspace*{-0.1in}}
\def\bml{\begin{mathletters}}
\def\eml{\end{mathletters}}
\def\beq{\begin{equation}}
\def\eeq{\end{equation}}
\def\bea{\begin{eqnarray}}
\def\eea{\end{eqnarray}}
\def\ba{\begin{array}}
\def\ea{\end{array}}

\def\nn{\nonumber}
\def\to{\rightarrow}
\def\e{{\rm e}}

\def\d{\delta}
\def\L{\Lambda}
\def\l{\lambda}
\def\o{\omega}
\def\z{\bar{z}}
\def\Z{\bar{Z}}
\def\tr{\,{\rm tr}\,}
\def\det{\,{\rm det}\,}
\def\Hd{H^{\dagger}}
\def\Hb{\bar{H}}
\def\Qd{Q^{\dagger}}

\def\Qt{\tilde{Q}}

\def\Ui{U^{-1}}

\def\phb{\bar{\phi}}
\def\psb{\bar{\psi}}
\def\chb{\bar{\chi}}
\def\df{d\psb d\psi d\chb d\chi}

\def\dbb{d\phb d\phi}
\def\s{\sqrt{s}}

\def\C{{\bf C}}
\def\R{{\bf R}}
\def\Q{{\bf H}}
\begin{document}
\preprint{}
\draft
\title{
Replica treatment of non-Hermitian disordered Hamiltonians}
\author{Shinsuke M. Nishigaki\ and\  Alex Kamenev}
\address{
Department of Physics, Technion--Israel Institute of Technology,
Haifa 32000, Israel
}
\date{September 13, 2001}
\maketitle
\begin{abstract}
We employ the fermionic and bosonic replicated
nonlinear $\sigma$ models to treat
Ginibre unitary, symplectic, and orthogonal ensembles of
non-Hermitian random matrix Hamiltonians.
Using saddle point approach combined with Borel resummation procedure 
we derive the exact large--$N$ results for 
microscopic 
density of states in all 
three ensembles.
We also obtain tails of the density of states as well the two-point 
function for the unitary ensemble.
\end{abstract}
\setcounter{footnote}{0}
\narrowtext

\section{Introduction}
\label{s1}  
In recent years the non-Hermitian random matrix Hamiltonians found a 
broad
spectrum of applications in physical problems\cite{FyoR}.
Their applicability ranges from dynamics of vortex lines\cite{HN}
and gauge theory at high baryon density\cite{Ste}
to description of biological populations\cite{NS}.
On the mathematical side, the study of non-Hermitian ensembles was
pioneered by Ginibre\cite{Gin},
who introduced the non-Hermitian counterparts of the Dyson ensembles
and computed correlation functions for the unitary ensemble.
A few decades later Mehta extended
his results to the symplectic ensemble\cite{Meh}.  Feinberg and
Zee\cite{FZ} introduced the Hermitization trick and derived a number
of results going beyond Ginibre Gaussian ensembles.  The
field-theoretical treatment\cite{Weg} in the supersymmetric
formulation\cite{Efe83} was introduced by Fyodorov, Sommers, and
Khoruzhenko\cite{FKS} for the unitary ensemble.  Efetov\cite{Efe,KE}
has subsequently developed the supersymmetric method for all three
non-Hermitian ensembles.  In this paper we take an alternative route
of the field-theoretical treatment, namely the replica trick\cite{EA}. 
This method was recently proved to be capable of describing the
correlation functions of the Hermitian Dyson ensembles
\cite{KM1,KM2,YL,DV,ADDV}. 
So far the validity of
the replica approach has been
restricted to the asymptotic regions of the
correlation functions only, while the supersymmetry is exact
everywhere.  It has, however, an important advantage of being
applicable to much broader class of systems, e.g.\ interacting
disordered systems\cite{Fin} or Calogero--Sutherland models\cite{GK}.

Here we show that the replica strategy works for the Ginibre 
ensembles as
well.
Although the treatment is philosophically similar to that for the
Dyson ensembles, 
it requires considerable technical modifications associated with the
intrinsic
chirality of the non-Hermitian symmetry classes.
In particular, the Riemannian symmetric manifolds which are target 
spaces of
the nonlinear $\sigma$ model
differs from
the corresponding Hermitian counterparts. 
Rather, they are 
subspaces of
their chiral Hermitian counterparts, as may be anticipated upon
Hermitization.  This necessitates introduction of convenient
parameterizations, calculation of the measure and finally an
appropriate analytical continuation.  
We also extend the method by performing all order expansion around
the saddle points.  
In the $n\to 0$ limit the resulting series may be
summed up employing the Borel resummation procedure.  As a result, we
are able to obtain formally exact results for the density of states
for all three Ginibre ensembles.

The paper is organized as follows: in Section \ref{s2} we collect
analytic results known for Ginibre ensembles, and preliminaries about 
the
replica method.  In Section \ref{s3} we derive
unitary
$\sigma$ model using fermionic replica and obtain corresponding
density of states, including the tails near the boundary of the
spectrum support.  Section \ref{s4} extends the unitary $\sigma$ model
to calculation of the two-point function.  In Section \ref{s5} we
treat the density of states for the symplectic ensemble using
fermionic replica.  In Section \ref{s6} we treat the density of states
for the orthogonal ensembles using bosonic replica.  Finally
peculiarities of the method and possible extensions are discussed in
Section \ref{s7}.

\section{Preliminaries}
\label{s2}
\subsection{Non-Hermitian Gaussian ensembles}
Throughout this paper we adopt Mehta's conventions and
notations\cite{Meh}.
Ginibre ensemble with 
$\beta=1$ (orthogonal), $\beta=2$ (unitary), or
$\beta=4$ (symplectic)
is defined as  ensembles of $N\times N$
matrices $H$ over one of the fields:
${\bf R}$,  ${\bf C}$, or ${\bf H}$. Each entry of such matrices 
contains
$\beta$ real components which 
are
i.i.d.\ variables
drawn from  the Gaussian distribution \cite{Gin}
\beq
d\mu(H)=\left({\alpha}{\pi}\right)^{-\beta N^2/2}\,
\e^{-\frac{1}{\alpha}\tr H H^\dagger}
\prod_{q=1}^\beta \prod_{i,j=1}^N  dH^{(q)}_{ij}
\eeq
($\alpha=2,1,1$ for $\beta=1,2, 4$).

For a real $N\times N$ matrix $H$ ($\beta=1$), the secular equation, 
$\det(z-H)=0$,  has real coefficients, so
complex eigenvalues always appear in conjugate pairs.
For a quaternion-real $N\times N$ matrix ${H}$ ($\beta=4$),
one can find a complex $2N\times 2N$ matrix representation 
(also denoted as $H$ for notational simplicity)
satisfying
\beq
\bar{H}=J H J^{-1}, \ \
J=\left[
\ba{cc}
0 & \openone_N\\
-\openone_N & 0
\ea
\right].
\label{JHJ}
\eeq
It implies that
$\det(\bar{z}-{H})=\overline{\det({z}-{H} )},$
and thus complex eigenvalues again appear in conjugate pairs.

Spectral correlation functions are defined as
\beq
R_p(z_1,\ldots,z_p)=\langle\tr \delta^2(z_1-H) \cdots \tr 
\delta^2(z_p-H)
\rangle,
\eeq
where $\langle\cdots\rangle$ denotes averaging with $d\mu(H)$.
The macroscopic density of states (DoS) of $H$ is 
uniform within a circle of radius $\sqrt{N}$
($\beta=1,2$)
or $\sqrt{2N}$ ($\beta=4$).
Below we collect known results for the microscopic correlation
functions
computed by the (skew-)orthogonal
polynomial method\cite{Gin,Meh} and by the supersymmetry
method\cite{Efe}.\\

\noindent
\underline{Unitary Ensemble}\cite{Gin,Meh,fn1}
\bml
\bea
&&\pi R_1(z)=\frac{\e^{-|z|^2} }{\Gamma(N)}
\int\limits_{0}^\infty d\lambda\,\e^{-\lambda} (\lambda+|z|^2)^{N-1}
\ \ \ \mbox{(exact)},
\label{DoS2fin}\\
&&\pi R_1(z)
=1\ \ \ (|z|\ll \sqrt{N}),
\ \ \ 
=0\ \ \ (|z|\gg \sqrt{N}),
\label{DoS2}
\\
&&\pi R_1(z)=
1-\frac{\e^{-2u^2}}{2\sqrt{2\pi}u}
\ \ \ (|z|=\sqrt{N}- u,\ 1\ll u \ll \sqrt{N}),
\nn\\
&&
\label{DoS2e}\\
&&\pi^2 R_2(z,z')=1-\e^{-|z-z'|^2} .\ \ \
(|z|,|z'| \ll \sqrt{N})
\label{TPF2}
\eea
\label{DoSTPF2}
\eml
$\!\!\!\!$
\underline{Symplectic Ensemble}\cite{Meh,Efe}
\ \ \ $(z=x+i y,\ |z| \ll \sqrt{N})$
\beq 
{\pi}R_1(z)=
2y^2\int\limits_0^1 \frac{d\lambda}{\sqrt{1-\lambda}}\e^{-2y^2 \lambda}.
~~~~~~~~~~~~~~
\label{DoS4}
\eeq
\underline{Orthogonal Ensemble}\cite{Efe}
\ \ \ $(z=x+i y,\ |z| \ll \sqrt{N})$
\beq 
{\pi}R_1(z)=
2y^2\int\limits_0^\infty \frac{d\lambda}{\sqrt{1+\lambda}}\e^{-2y^2 \lambda}
+\sqrt{\pi}\delta(y).
\label{DoS1}
\eeq

\subsection{Replica method}
In the Hermitian case, one uses the identity 
\bea
\delta(x)&=&\frac{1}{\pi}{\rm Im} \frac{1}{x-i \epsilon}
=\frac{1}{\pi}{\rm Im} \frac{d}{dx}\log(x-i \epsilon) \nn\\
&=&\lim_{n\to 0}\frac{1}{\pi n}{\rm Im} \frac{d}{dx} (x-i\epsilon)^n,
\nn 
\eea 
which in the matrix context takes the form\cite{EA} \beq
\tr\delta (x-H)= \lim_{n\to 0}\frac{1}{\pi n} {\rm Im}\frac{d}{dx}
{\rm det}^n(x-i\epsilon-H) .  \eeq The two distinct versions $n > 0$
and $n < 0$ are called fermionic and bosonic replica, correspondingly. 
The fermionic replicated generating functions (${\rm det}^n(x-H)$
averaged with $d\mu(H)$) of unitary \cite{KM1}, orthogonal, symplectic
\cite{KM2,YL}, and chiral unitary \cite{DV} ensembles lead to compact
nonlinear $\sigma$ models on ${\rm U}(2n)/{\rm U}(n)\times {\rm
U}(n)$, ${\rm Sp}(2n)/{\rm Sp}(n)\times {\rm Sp}(n)$, ${\rm
O}(2n)/{\rm O}(n)\times {\rm O}(n)$, and ${\rm U}(2n)$, respectively,
which are FF blocks of the Riemannian symmetric superspaces of type
AIII$|$AIII, BDI$|$CII, CII$|$BDI, and A$|$A\cite{Zir}.  After
appropriate parameterization of these symmetric spaces one obtains
$n$-fold compact integrals, which may be evaluated in the saddle point
approximation.  By collecting contributions from dominant as well as
subdominant saddle points, one obtains the asymptotic expressions for
the correlation functions.  Bosonic replica has been successfully
applied to the chiral unitary ensemble\cite{DV}, see also
Ref.\cite{Fyo}.

In the non-Hermitian case, one may employ the  identity
\[ 
\delta^2(z)=\frac{1}{\pi}\partial_z\partial_{\z} \log (z\z)
=\lim_{n\to 0}\frac{1}{\pi n} \partial_z\partial_{\z}(z\z)^n, 
\] 
which in our context takes the form 
\beq \tr\delta^2(z-H)
=\lim_{n\to 0} \frac{1}{\pi n}\partial_z\partial_{\z} {\rm det}^n(z-H)
{\rm det}^n (\z-H^\dagger).
\label{d2}
\eeq
This form resembles the generating function for a block
off-diagonal (chiral) matrix obtained upon Hermitization\cite{Has}. 
The difference is that the spectral parameter enters off-diagonally
rather than diagonally as in the chiral case.  This distinction
however does not affect the resulting saddle point manifolds, which is
determined at infinitesimal spectral parameters.  Therefore, according
to Zirnbauer's celebrated result\cite{Zir}, the saddle point manifolds
are expected to be (subspaces of) ${\rm U}(2n)$, ${\rm U}(2n)/{\rm
O}(2n)$, ${\rm U}(2n)/{\rm Sp}(n)$ (for fermionic replica) or ${\rm
GL}(2n,\C)/{\rm U}(2n)$, ${\rm U}^*(2n)/{\rm Sp}(n)$, ${\rm
GL}(2n,\R)/{\rm O}(2n)$ (for bosonic replica) for $\beta=2,4,1$,
respectively, which are FF and BB blocks of the Riemannian symmetric
superspaces of type A$|$A, AII$|$AI, and AI$|$AII.

Our aim is to obtain the DoS and the two-point function
(cf.\ Eqs.\ (\ref{DoSTPF2}), (\ref{DoS4}), (\ref{DoS1}))
including exponential terms, using the replica strategy.  We employ
either fermionic or bosonic replica, whichever is suitable for the
actual computation.

\section{Ginibre unitary ensemble: \\ Density of States}
\label{s3}
Consider the fermionic replicated generating function (positive
moment of the characteristic polynomial) of Ginibre unitary ensemble
defined as 
\beq Z_n(z,\z)= \int_{\C^{N \times
N}}\!\!\!\!\!\!\!dH\,\e^{-\tr H \Hd} \det^n(z-H) \det^n(\z-\Hd), \eeq
where $dH=\pi^{-N^2}\prod_{i,j}^N d^2 H^{ij}.$ We denote `color'
indices by $i,j=1,\ldots,N$ and `flavor' indices by $a,b=1,\ldots,n$. 
In the standard way one introduces mutually independent Grassmann
variables $\psi_{a}^i, \psb_{a}^i, \chi_{a}^i, \chb_{a}^i$, and
rewrites the generating function in terms of an auxiliary matrix
$Q_{ab}$ (a well-known reexpression of the chiral unitary ensemble
with flavor and color indices interchanged\cite{SV,AST}):
\widetext
\bea
Z_n(z,\z)
&=&\int_{\C^{N \times N}}\!\!\!\!\!\!\!dH\,\int\df\,\exp\left(
-H^{ij} \Hb^{ij}
-\psb_{a}^i(z \delta^{ij}-H^{ij})\psi_{a}^j
-\chb_{a}^j(\z \delta^{ji}-\Hb^{ij})\chi_{a}^i \right)\nn \\
&=&\int\df\,\exp(
\psb_{a}^i\psi_{a}^j\chb_{b}^j\chi_{b}^i
-z \psb_{a}^i\psi_{a}^i-\z \chb_{a}^i\chi_{a}^i
)\nn\\
&=&\int_{\C^{n \times n}}\!\!\!\!\!\!dQ\,\e^{-\tr Q \Qd}
\int\df\,
\exp\left( -
[\psb_a^i \ \chb_a^i]
\left[
\ba{cc}
z \delta_{ab} & -Q_{ab} \\
\Qd_{ab}     &  \z \delta_{ab}
\ea
\right]
\left[
\ba{c}
\psi_b^i \\
\chi_b^i
\ea
\right]
\right)\nn\\
&=&\int_{\C^{n \times n}}\!\!\!\!\!\!dQ\,\e^{-\tr Q \Qd}
\det^N \left[
\ba{cc}
z & -Q \\
\Qd &  \z
\ea
\right]
=\int_{\C^{n \times n}}\!\!\!\!\!\!dQ\,\e^{-\tr Q \Qd}
\det^N\left( z\z + Q \Qd \right).
\eea
\Rrule
\narrowtext
\noindent
Here $d\psi=\prod_{a,i}d\psi^i_{a}$ and so forth, and
$dQ=\pi^{-n^2}\prod_{a,b}^n d^2 Q_{ab}$.
A complex matrix $Q\in {\C^{n \times n}}$
can be uniquely written as (singular value decomposition)
\bea
&&Q=U \L V,\ \ \ U\in {\rm U}(n)/{\rm U}(1)^n,\ V \in {\rm U}(n),
\nn\\
&&\L ={\rm diag}({\lambda_a^{1/2}}) ,\ \lambda_a\geq 0,
\eea
and a Euclidean measure $dQ$ on $\C^{n \times n}$ is related to
normalized Haar measures $dU$ on ${\rm U}(n)/{\rm U}(1)^n$ and $dV$ on
${\rm U}(n)$ by \beq dQ=dU\,dV\,\Delta(\lambda)^2\prod_{a=1}^n
d\lambda_a , \ \ \Delta(\lambda)=\prod_{a>b}^n (\lambda_a-\lambda_b).
\eeq As the integrand does not depend on $U$ and $V$,
$Z_n(z,\z)$ is written as an $n$-fold integral
(even for finite $N$),
\beq
Z_n(z,\z)=\int\limits_0^\infty
\prod_{a=1}^n \left(
d\lambda_a\, \e^{-\lambda_a} (\lambda_a+|z|^2)^N \right)
\Delta(\lambda)^2 ,
\label{z2i}
\eeq
up to an irrelevant constant factor that approaches unity
in the replica limit.
Notice the striking resemblance
between the DoS (\ref{DoS2fin}) and the replica generating function
(\ref{z2i}) even at {\em finite} $N$.
This raises the question of whether the replica method
may be applicable beyond the asymptotic analysis.
We come back to this issue in Sect.\ref{s7}.

\subsection{DoS in the bulk}
Now we take the large--$N$ limit.
For $|z|<\sqrt{N}$, the saddle point equation
\beq
1-\frac{N}{\lambda+|z|^2}=0
\eeq
has a solution $\lambda=N-|z|^2$. Therefore
\beq
Z_n(z,\z)=
{}\e^{-n (N-|z|^2)}.
\eeq
up to an irrelevant constant given by the Selberg integral\cite{Meh},
which goes
to unity as $n\to 0$, and ${\rm O}(n^2)$ terms.
On the other hand, for $|z|>\sqrt{N}$,
the integral is dominated by the end point $\lambda=0$:
\beq
Z_n(z,\z)=
|z|^{2nN}.
\eeq
Using  relation (\ref{d2}), one obtains for 
the DoS 
\beq
\pi R_1(z)=
\lim_{n\to 0}\frac{1}{n}\partial_z\partial_{\z} {Z_n}(z,\z)
=
\left\{
\ba{ll}
1\ \ & |z|<\sqrt{N} \\
0 \ \ & |z|>\sqrt{N}
\ea
\right. .
\eeq
Therefore $N$ complex eigenvalues are uniformly distributed within
a circle of radius $\sqrt{N}$, in agreement with the exact result
(\ref{DoS2}).

\subsection{DoS at the edge}
If $|z|$ is close to the edge of the  circle, such that $|z|=\sqrt{N} 
-u$,
where 
$1{< \atop \sim} u \ll \sqrt{N}$, one may work out corrections to the
uniform DoS. 
In this case both the saddle point at
$\lambda_a = N-|z|^2\simeq 2\sqrt{N}u$ and the end point $\lambda_a = 
0$
contribute to the partition function.  Summing up all the
contributions one finds 
\bea
                                             \label{edge}
&&Z_n(z,\bar z) = \e^{-n(N-|z|^2)} \sum\limits_{p=0}^n {n\choose 
p}(-)^p
(2\sqrt{N}u)^{2p(n-p)} \e^{-2p u^2} \nonumber \\
&& \int\limits_0^{\infty} \prod\limits_{a=1}^p \left( d\lambda_a\,
\e^{-{2u\over \sqrt{N} } \lambda_a} \right) \Delta(\lambda)^2
\times
\int\limits_{-\infty}^{\infty} \prod\limits_{b=1}^{n-p} \Bigl( 
d\lambda_b \,
\e^{-{\lambda_b^2 \over 2N} } \Bigr) \Delta(\lambda)^2\, . \nonumber\\
\eea
Employing  Selberg integrals and neglecting 
constant factors which go to 
unity in
the 
$n\to 0$ limit, one obtains
\bea
                                             \label{edge1}
&&Z_n(z,\bar z) =\\
&&\e^{-n(N-|z|^2)}
\sum\limits_{p=0}^n  (-)^p \e^{-2p u^2} (2u)^{2p(n-p)-p^2} 
(2\pi)^{n-p\over
2}  D^p_n \, ,
\nonumber
\eea
where 
\beq 
D^p_n\equiv  \prod\limits_{a=1}^p
\frac{ \Gamma(a)^2}{\Gamma(n-a+1)} \, .
                                           \label{edge2}
\eeq
Since $D^p_n =0$ for $p>n$, one may extend  summation over $p$ to 
infinity in
Eq.~(\ref{edge1}). We then perform  the analytic continuation $n\to 0$.
To this end we notice that $D^p_n = O(n^p)$ in the limit $n\to 0$.
Therefore only the terms with $p=0$ and $p=1$ contribute to the 
partition
function in the small--$n$ limit.
As a result, one finds
\beq
                                             \label{edge3}
Z_n(z,\bar z) = \e^{-n(N-|z|^2)}
\left( 1- n \frac{\e^{-2u^2}}{\sqrt{2\pi}(2u)^{3}} \right) .
\eeq
Employing finally Eq.~(\ref{d2}) one obtains for the DoS close to the
edge of the spectrum support
\bea 
\pi R_1(z) &=&\lim_{n\to 0}\frac{1}{n}\partial_z \partial_{\z}
Z_n(z,\z) \nn\\
&=&
1 - \frac{\e^{-2u^2}}{2\,{\sqrt{2\pi }}u} +
  {O}\left(\frac{\e^{-2u^2}}{u^3}\right)\, .
\eea
This asymptotics agree with the exact result (\ref{DoS2e})
within the validity of our approximation
$1\ll u^2 (\ll N)$. 
The tail of the DoS outside the circle,
$|z|>\sqrt{N}$ can not be obtained in the fermionic replica.
Indeed, in this case the saddle point is situated at negative 
$\lambda$ and 
the contour of integration 
cannot be deformed to pass through it. 
As a result the 
integrals are dominated by the end point $\lambda = 0$, which leads 
to 
the zero DoS for $|z|>\sqrt{N}$. It is possible that the bosonic 
replica 
$\sigma$ model may be capable to produce these tails 
(c.f. Ref.\cite{CGP}).

\section{Ginibre unitary ensemble: two-level correlation}
\label{s4}
Consider the fermionic replicated generating function
with two spectral parameters,
\bea
&&Z_{n}(z_1,\z_1;z_2,\z_2)
=
\int_{\C^{N \times N}}\!\!\!\!\!\!\!dH\,\e^{-\tr H \Hd}
\times
\\
&&
\det^{n}(z_1\!-\!H) \det^{n}(\z_1\!-\!\Hd)
\det^{n}(z_2\!-\!H) \det^{n}(\z_2\!-\!\Hd).
\nn
\eea
Introducing the enlarged flavor indices
$A, B=1,\ldots,2n$ and a $2n\times 2n$ matrix $Z$,
\[
Z=
\left[
\ba{cc}
z_1 \openone_n & 0\\
0 & z_2 \openone_{n}
\ea
\right] \, ,
\]
one rewrites the generating function in terms of
the  auxiliary matrix $Q_{AB}$ 
as in the previous section,
\beq
Z_{n}(z_1,\z_1;z_2,\z_2)
=
\int_{\C^{2n \times 2n}}\!\!\!\!\!\!\!\!\!\!dQ\,\e^{-\tr Q \Qd}
\det^N \left[
\ba{cc}
Z & -Q \\
\Qd & \Z
\ea
\right]  ,
\eeq
with $dQ=\pi^{-4n^2}\prod_{A,B}^{2n} d^2 Q_{AB}$.

Hereafter we concentrate on the center of the circle
$|z_1|, |z_2| \ll \sqrt{N}$.
At $z_1=z_2=0$, 
the large--$N$ saddle point equation
\beq
\Qd-N\Qd (Q \Qd)^{-1}=0 ,
\eeq 
is solved by 
\beq Q =\sqrt{N} U, \
\ U\in {\rm U}(2n).
\eeq 
For $z_1, z_2$ finite and of order unity,
one can write for such $Q$,
\bea
&& \det \left[ \ba{cc} Z & -Q \\
\Qd & \Z
\ea
\right]
\simeq
{}
\exp
\left(
- \frac1{2N}\,{\rm tr}
\left[
\ba{cc}
0 & -\Ui\Z\\
U Z & 0
\ea
\right]^2
\right)
\nn\\
&&=
{}\exp \left(
\frac{1}{N}
\left(
2n\left| z \right|^2 +
\left| \frac{\omega}{2} \right|^2
\tr U s \Ui s  \right) \right),
\eea
where
\[
z=\frac{z_1+z_2}{2} ,\ \ \omega={z_1-z_2},\ \
s=
\left[
\ba{cc}\openone_n & 0 \\ 0 & -\openone_n \ea \right].
\]
Therefore,
\bea 
&&
Z_{n}(z_1,\z_1;z_2,\z_2)=
\nn\\
&&
{}\e^{-2n(N-\left| z \right|^2)}
\int_{{\rm U}(2n)}\!\!\!\!\!\!\!dU\,
\exp\left(
\frac{|\omega|^2}{4}
\tr U s \Ui s
\right),
\label{UUU}
\eea
up to an irrelevant constant factor that approaches unity
in the replica limit.
Here $dU$ is a Haar measure on ${\rm U}(2n)$.
The above integrand is invariant under
\[
U\to U
\left[
\ba{cc}
u & 0\\
0 & u'
\ea
\right],\ \ u, u'\in {\rm U}(n),
\]
so the saddle point manifold shrinks down to
${{\rm U}(2n)/{\rm U}(n)\times {\rm U}(n)}$.
Adopting the parameterization of this coset manifold employed by
Verbaarschot and Zirnbauer\cite{VZ}, 
\bea 
&& U s U^{-1} = \left[
\ba{cc} u & 0\\
0 & v
\ea
\right]
\left[
\ba{cc}
\cos {\bf \Theta} & \sin {\bf \Theta}\, \e^{i{\bf \Phi}}\\
\sin {\bf \Theta} \,\e^{-i{\bf \Phi}} & -\cos {\bf \Theta}
\ea
\right]
\left[
\ba{cc}
u^{-1} & 0\\
0 & v^{-1}
\ea
\right] ,
\nn\\
&&
{\bf\Theta}={\rm diag}(\theta_a),\
{\bf\Phi}={\rm diag}(\phi_a),\
\ \
u,v \in {\rm U}(n)/{\rm U}(1)^n,
\\
&&
dU=du\,dv\,\prod_{a=1}^n d\phi_a\,
\prod_{a=1}^n d\cos \theta_a\,\Delta(\cos \theta)^2,
\eea
and substituting
\beq
\tr U s U^{-1} s
=2\tr\cos {\bf \Theta}=  2\sum_{a=1}^n \cos \theta_a,
\eeq
one obtains
\bea
&&
Z_{n}(z_1,\z_1;z_2,\z_2)=
\nn\\
&&
{} \e^{-2n(N-|z|^2)}
\int\limits_{-1}^1 \prod_{a=1}^n
\left(
d\lambda_a\,
\e^{
\frac{\left| {\omega} \right|^2}{2}
\lambda_a}
\right)
\Delta(\lambda)^2 ,
\label{z2ii}
\eea 
where $\lambda_a=\cos \theta_a$.

The replica limit of the above $n$-fold integral
has been discussed extensively by
many authors\cite{VZ,YL,ADDV,GK}
(see also Ref.\cite{For} for its asymptotic analysis).  
The integral
(\ref{UUU}) over Grassmann manifold ${\rm U}(2n)/{\rm U}(n)\times{\rm
U}(n)$ with a height function $\tr U s U^{-1}s$ satisfies the criteria
of the Duistermaat--Heckman localization theorem\cite{DH,Zir99} and
thus the saddle point method applied to Eq.\ (\ref{UUU}) is exact for
an arbitrary value of $|\omega|^2$.  For positive large
$t=|\omega|^2/2$, the integral (\ref{z2ii}) may be `approximated' by
contributions from the end points $\lambda_a=\pm 1$, 
\bea 
&&
\int\limits_{-1}^1 \prod_{a=1}^n\left( d\lambda_a\, \e^{t \l_a}
\right)\, \Delta(\lambda)^2 
\nn\\
&& 
=\sum_{p=0}^n (-)^{p} (F^{p}_n(1))^2
\frac{\e^{(n-2p)t}}{(2t)^{(n-p)^2+p^2}}\, ,
\eea 
where 
\beq
F^p_n ({k}) \equiv {n\choose p}
\prod\limits_{a=1}^p   \frac{\Gamma(1+a/{k})}
{\Gamma\Big(1+(n-a+1)/{k}\Big)} \,
\label{fpn}
\eeq
for $p>0$ and $F^0_n ({k})=1$.
One may extend summation over $p$ to infinity, since the binomial
coefficient ${n\choose p}\equiv 0$ for $p>n$. After this one may 
perform
analytical continuation, employing the small--$n$ limit,
\beq
F_n^p({k}) = n {(-1)^{p+1} \over p} \prod\limits_{a=1}^p {
\Gamma(1+a/{k}) \over \Gamma(1-(a-1)/{k})} + {\rm O}(n^2)\, .
\label{Fn}  
\eeq  
Accordingly the integral reads, for small $n$,
\beq 
\int\limits_{-1}^1 \prod_{a=1}^n\left( d\lambda_a\, \e^{t \l_a}
\right)\, \Delta(\lambda)^2 \simeq \frac{\e^{nt}}{(2t)^{n^2}} -n^2
\frac{\e^{(n-2)t}}{(2t)^{n^2-2n+2}}\, .  
\eeq 
Using this formula, one
obtains 
\bea 
&&{Z_n}(z_1,\z_1;z_2,\z_2)\simeq \nn\\
&&{} \e^{-2n(N-|z|^2)}
\left(
\frac{\e^{n |\omega|^2/2}}{|\omega|^{2n^2}}-
n^2 \frac{\e^{(n-2) |\omega|^2/2}}{|\omega|^{2(n^2-2n+2)}}
\right).
\eea
The two-point function
is given by 
\bea 
\pi^2 R_2(z_1,z_2)&=&
\lim_{n\to 0}\frac{1}{n^2}
\partial_{z_1}\partial_{\z_1}\partial_{z_2}\partial_{\z_2}
{Z_n}(z_1,\z_1;z_2,\z_2) \nn\\
&=&
1-\e^{-|z_1-z_2|^2}
+O\left(
\frac{\e^{-|z_1-z_2|^2}}{|z_1-z_2|^2}
\right).
\eea
This agrees with the exact result (\ref{TPF2})
within the validity of approximation
$1\ll |z_1-z_2|^2 (\ll N)$.
The disconnected part $1=\pi^2 R_1(z_1)R_1(z_2)$ is given by the
replica symmetric saddle point $p=0$, whereas the connected part
$-{\e^{-|z_1-z_2|^2}}$ is given by the replica nonsymmetric saddle
point, $p=1$. 
In the latter case, the 
$(n-1)$-fold integral around $\l=1$ and
a one-fold integral around $\l=-1$ `interact' through the Van der 
Monde
determinant.  
Taking this interaction into account in the perturbative manner, 
one finds that all terms proportional to
${\e^{-|z_1-z_2|^2}}/{|z_1-z_2|^{2k}}$ ($k\geq 1$)
are canceled. 
We have checked this statement explicitly to the order $k=1$.
In general it is a  manifestation  of  the
Duistermaat--Heckman localization theorem\cite{Zir99}.

\section{Ginibre symplectic ensemble}
\label{s5}
For a $2N\times 2N$ complex matrix representation of
an $N\times N$ quaternion--real matrix $H$,
the constraint (\ref{JHJ})
may be solved by
\beq
H=
\left[
\ba{cc}
S & T\\
-\bar{T} & \bar{S}
\ea
\right], \ \ S, T \in \C^{N\times N}.
\eeq
The fermionic replicated generating function
of Ginibre symplectic ensemble thus reads
\widetext
\Lrule
\bea
Z_{n}(z,\z)
&=&\int_{\Q^{N \times N}}\!\!\!\!\!\!d{H}\,
\e^{-\frac12 \tr {H}{H}^\dagger}
\det^{n}(z-{H}) \det^{n}(\z-{H})
\nn\\
&=&\int_{\C^{N \times N}}\!\!\!\!\!\!\!dS\,dT\,
\e^{-\tr SS^\dagger-\tr TT^\dagger}
\det^{n}\left(z-
\left[
\ba{cc}
S & T\\
-\bar{T} & \bar{S}
\ea
\right]\right)
\det^{n}\left(\z-
\left[
\ba{cc}
S & T\\
-\bar{T} & \bar{S}
\ea
\right]\right),
\eea
where
$dH=dS dT=\pi^{-2N^2}\prod_{i,j}^N d^2 S^{ij}d^2 T^{ij}.$  Once again 
we
introduce enlarged flavor indices $A, B=1,\ldots,2n$ 
and express the generating function in terms of
the  auxiliary matrix $Q_{AB}$,
\bea
&&Z_{n}(z,\z)
=\int_{\C^{N \times N}}\!\!\!\!\!\!\!dS\,dT\!\!
\int\df
\exp\left(
-S^{ij} \bar{S}^{ij}-T^{ij} \bar{T}^{ij}
-[\psb_{A}^i\ \chb_{A}^i]
\left(
\left[
\ba{cc}
\delta^{ij} & 0\\
0 & \delta^{ij}
\ea
\right]Z_{AB}
-
  \left[
\ba{cc}
S^{ij} & T^{ij}\\
-\bar{T}^{ij} & \bar{S}^{ij}
\ea
\right] \delta_{AB}
\right)
\left[
\ba{c}
\psi_B^j\\
\chi_B^j
\ea
\right]
\right)
\nn\\
&&=
\int\df\exp\left(
-\psb_{A}^i Z_{AB} \psi_{B}^i
-\chb_{A}^i Z_{AB} \chi_{B}^i
-\psi_{A}^i \psb_{A}^j \chb_{B}^j \chi_B^i
+\chi_{A}^i \psb_{A}^j \chb_{B}^j \psi_B^i
\right)
\nn\\
&&=
\int_{{\cal M}}dQ
\int\df
\nn\\
&&
\ \ \ \ \exp\left(
-\frac12 Q_{AB} \bar{Q}_{AB}
+\frac12  Q_{AB} (\psb_A^i \chb_B^i + \psb_B^i \chb_A^i)
+\frac12  \bar{Q}_{BA}  (\chi_A^j \psi_B^j + \chi_B^j \psi_A^j)
-\psb_{A}^i Z_{AB} \psi_{B}^i
-\chb_{A}^i Z_{AB} \chi_{B}^i
\right)
\nn\\
&&=
\int_{{\cal M}}dQ\,
\e^{-\frac12 \tr Q Q^\dagger}
\int\df\exp\left( -
[\psb_A^i\ \chi_A^i]
\left[
\ba{cc}
Q_{AB} & -Z_{AB}\\
Z_{AB} & Q^\dagger_{AB}
\ea
\right]
\left[
\ba{c}
\chb_B^i\\
\psi_B^i
\ea
\right]
\right)
=
\int_{{\cal M}}dQ\,
\e^{-\frac12 \tr Q Q^\dagger}
\det^N
\left[
\ba{cc}
Q& -Z\\
Z& Q^\dagger
\ea
\right] ,
\eea
\narrowtext
\noindent
where $z=x+i y,\ 
Z=x\openone_{2n} +i y s .$ 
Here $dQ=(2\pi)^{-2n^2-n}\prod_{A\geq B}^{2n} d^2 Q_{AB}$,
and the integration domain
${\cal M}$ is a set of $2n\times 2n$  complex symmetric matrices,
because its antisymmetric part decouples from the fermionic bilinear.
A similar procedure was adopted for the chiral symplectic
ensemble\cite{HV}.

Hereafter we concentrate on the center of the circle $|z|\ll
\sqrt{2N}$.
The large--$N$ saddle point equation
\beq
\frac12 \Qd-N\Qd (Q \Qd)^{-1}=0 ,
\eeq
is solved by
\beq
Q =\sqrt{2N}U,\ \  U\in {\rm U}(2n).
\eeq
For $z$ finite and of order unity,
one can write for such $Q$,
\bea
&&\det
\left[
\ba{cc}
Q & -Z \\
Z & \Qd
\ea
\right]
\simeq
{}
\exp 
\left(
-\frac{1}{4N}
\, {\rm tr}
\left[
\ba{cc}
0 & -\Ui Z\\
U Z & 0
\ea
\right]^2
\right)
\nn\\
&&=
{}
\exp \left( \frac{1}{N}
\left( n x^2 -\frac{y^2}{2} \tr U s \Ui s  \right) \right)\, .
\eea
Therefore,
\beq
Z_{n}(z,\z)=
\e^{-n(2N- x^2)}
\int_{{{\rm U}(2n)\atop U=U^T}}\!\!\!dU\,
\exp\left(
-\frac{y^2}{2}
\tr U s \Ui s
\right),
\label{UUT}
\eeq
up to an irrelevant constant factor that approaches unity
in the replica limit.
Here $dU$ is a 
Haar measure on a group quotient ${\rm U}(2n)/{\rm O}(2n)$ 
that is
isomorphic to the space of $2n\times 2n$ unitary symmetric matrices
through a canonical projection $g\mapsto U=gg^T$.  Since the integrand
is invariant under \beq U\to \left[ \ba{cc} u & 0\\
0 & u'
\ea
\right]
U
\left[
\ba{cc}
u^T & 0\\
0 & u'{}^T
\ea
\right],\ \ u, u'\in {\rm U}(n),
\eeq
the saddle point manifold shrinks down to
the intersection of ${{\rm U}(2n)/{\rm O}(2n)}$ and
${{\rm U}(2n)/{\rm U}(n)\times {\rm U}(n)}$.
Following Zirnbauer and Haldane\cite{ZH},
one can parameterize the matrix $U$ and the measure $dU$ as
\bea U &=&
\left[ \ba{cc} u & 0\\
0 & v
\ea
\right]
\left[
\ba{cc}
\cos {\bf \Theta} & i\sin {\bf \Theta}   \\
i \sin {\bf \Theta}  & \cos {\bf \Theta}
\ea
\right]
\left[
\ba{cc}
u^T & 0\\
0 & v^T
\ea
\right] ,
\label{Usymm}\\
&&
{\bf \Theta}={\rm diag}(\theta_a),\
u,v \in {\rm U}(n),
\nn\\
dU&=&du\,dv\,
\prod_{a=1}^n  d\sin \theta_a\, |\Delta(\cos^2 \theta)|.
\label{dU}
\eea
Computation of the above Jacobian is detailed in Appendix A.
By substituting
\beq
\tr U s \Ui s
=4\sum_{a=1}^n \cos^2 \theta_a -2n,
\eeq
we obtain  the generating function
in the form of the $n$-fold integral
\beq
Z_{n}(z,\z)=
\e^{-n(2N-|z|^2)}
\int\limits_{0}^1\!\! \prod\limits_{a=1}^n
\left( \frac{d \lambda_a}{\sqrt{1-\lambda_a }} \,
\e^{-2 y^2 \lambda_a} \right)
|\Delta(\lambda)| \, ,
\label{Zn4}
\eeq
where $\lambda_a = \cos^2 \theta_a$.
Again we have encountered a striking resemblance
between the DoS (\ref{DoS4}) and the generating function
(\ref{Zn4}).

For $y^2 \gg 1$ this integral is dominated by  vicinities of the
two end points.
By considering $t\equiv 2y^2$ to have a
negative imaginary part (possibly infinitesimal),
one may deform the integration contour to the
upper half plane as
$[0,1]\to[0,i\infty)+(1+i\infty,1]$.
Employing  Selberg integrals, one may  evaluate
all the contributions with $p$ integrals
along $[1,1+i\infty)$
taken in the vicinity
of $\lambda=1$, whereas the remaining $n-p$ integrals
along $[0,i\infty)$ coming from
the vicinity of $\lambda =0$.
Whenever the end point $\lambda_{a} =0$ is adopted for
some $a$, 
one may safely expand $(1-\lambda_a)^{-1/2}$ in $\lambda_a$,
while it ought to be treated exactly when
the other end point $\lambda_{a} =1$ is chosen.
Meanwhile we approximate $(1-\lambda_a)^{-1/2}\simeq 1$ for the
former case and find
\bea
I_n(t)&=&
\int\limits_{0}^1 \prod\limits_{a=1}^n
\left( \frac{d \lambda_a}{\sqrt{1-\lambda_a }} \,
\e^{-t \l_a} \right)
|\Delta(\lambda)|
\nn\\
&\simeq&
\sum_{p=0}^n (F^{p}_n(2))^2
\frac{\Gamma\left({1+n-p\over 2} \right) }
{\Gamma\left({1\over 2} \right) }
\frac{(-1)^{p^2\over 2} \e^{-pt} }
{t^{{(n-p)(n-p+1)\over 2} + {p^2\over 2} } }\, .
\eea 
As in previous
sections, one may extend summation over $p$ to infinity and perform
analytical continuation, employing the small--$n$ limit (\ref{Fn}).
Only $p=0$ and $p=1$ terms contribute in the replica limit $n\to 0$,
since $p\geq 2$ terms are of the order ${\rm O}(n^2)$.  As a result,
\beq 
I_{n}(t)\simeq t^{-{n/ 2} } + n {i\sqrt{\pi} \over 2}t^{-{1/ 2}
}\e^{-t}.  
\eeq 
In order to obtain perturbative (i.e.\ powers of $1/t$) corrections,
one expand the $(1-\lambda_a)^{-1/2}$ factor in powers of $\lambda_a$
for the replica-symmetric ($p=0$) term, 
\bea
&&\int\limits_{0}^{i\infty} \prod\limits_{a=1}^n \left(d\lambda_a \,
\e^{-t \l_a} \right) |\Delta(\lambda)| \prod\limits_{b=1}^n
(1-\lambda_b)^{-1/2} \nn\\
&=& 
\int\limits_{0}^{i\infty} \prod\limits_{a=1}^n \left(d\lambda_a \,
\e^{-t \l_a} \right)
|\Delta(\lambda)| 
\exp \left(\sum_{k=1}^\infty \frac{\o_k}{2k}\right),
\label{wk}
\eea
where
$\o_k\equiv\sum_{a=1}^n \l_a^k$ are elemantary symmetric polynomials.
In Appendix B we prove by loop equations\cite{Mig}
that the RHS of  Eq.\ (\ref{wk}) is equal to
\bea
&&\int\limits_{0}^{i\infty} \prod\limits_{a=1}^n \left(d\lambda_a \,
\e^{-t \l_a} \right)
|\Delta(\lambda)| 
\exp \left(\sum_{k=1}^\infty \frac{\o_1^k}{2k}\right)
\label{w1}\\
&=&
\int\limits_{0}^{i\infty} 
\prod\limits_{a=1}^n
\left({d\lambda_a} \,\e^{-t \l_a} \right)|\Delta(\lambda)|
\Bigl(1-\sum_{b=1}^n\lambda_b \Bigr)^{-1/2}
\nn\\
&=&
(1+\partial_t)^{-1/2}
\int\limits_{0}^{i\infty} 
\prod\limits_{a=1}^n
\left({d\lambda_a} \,\e^{-t \l_a} \right)|\Delta(\lambda)|
\nn
\eea
plus terms of order $O(n^2)$.
Here
$(1+\partial_t)^{-1/2}$ stands for a formal series
\beq
(1+\partial_t)^{-1/2}=
\sum_{k=0}^\infty\frac{(2k-1)!!}{(-2)^k k!}\frac{d^k}{dt^k}.
\label{partial}
\eeq
Accordingly the generating function reads
\bea
Z_n(z,\z)&=&
{} \e^{-n(2N-|z|^2)}\,I_n(2y^2),\nn\\
I_n(t)&=&
(1+\partial_t)^{-1/2}t^{-{n/ 2} } + n {i\sqrt{\pi} \over 2}t^{-{1/ 2}
}\e^{-t}.  
\eea 
The DoS is then given by
\bea 
&&\pi R_1(z)=\lim_{n\to
0} \frac1n \partial_z\partial_{\z} Z_n(z,\z) \nn\\
&&
~~~~~~~
=
1-
\frac12 \{\partial_t ,t\}\partial_t 
(1+\partial_t)^{-1/2} \log t +i\sqrt{\pi t}\, \e^{-t} ,
\label{DoS4rep}
\eea
$t=2y^2$, up to terms of order $O(t^{-1/2}\e^{-t})$.
On the other hand, one can evaluate the asymptotic expansion
of the exact result (\ref{DoS4})
for ${\rm Re}\,y^2 \gg1$ and ${\rm Im}\,y^2<0$
by using the same saddle point method (it is formally equivalent to 
the 
calculation given  above for  $n=1$),
\bea
\pi R_1(z)&=&
t \left( \int\limits_{0}^{i\infty} \frac{d\lambda}{\sqrt{1-\lambda}} \,
\e^{-t \lambda} - \e^{-t } \int\limits_{0}^{i\infty}
\frac{d\lambda'}{\sqrt{-\lambda'}} \, \e^{-t \lambda'} \right) \nn\\
&=&
t(1+\partial_t)^{-1/2}t^{-1} +i\sqrt{\pi t}\,\e^{-t} .
\label{DoS4ex}
\eea
One may confirm the equality of Eqs.\ (\ref{DoS4rep}) and
(\ref{DoS4ex}) either by using a canonical commutation relation
$[\partial_t, t]=1$ to bring e.g.\ all $t$'s to the right of
$\partial_t$'s, or by explicit substitution of 
Eq.~(\ref{partial}) to derive an identical asymptotic series
\beq
                                   \label{borel}
\pi R_1(z)=
\sum_{k=0}^\infty \frac{(2k-1)!!}{(2y)^{2k}}
+i\sqrt{2\pi}y\, \e^{-2y^2} .
\eeq
The formal Borel resummation of the series, Eq.~(\ref{borel}), leads
the exact result for the DoS, Eq.~(\ref{DoS4}).
Therefore our fermionic replica method gives correctly
all orders of perturbative terms as well as
the 
nonperturbative term of the DoS. 
Although the imaginary terms $\pm i\sqrt{2\pi}y\, \e^{-2y^2}$
are artifacts of locating $y^2$ in the positive side of
the lower or upper half plane
and are obviously
absent for real positive $y^2$ (i.e.\ the positive real
axis is the Stokes
line for the asymptotic expansions),
it is their presence and coincidence that enables
us to identify the two expressions: 
(\ref{Zn4}) with a Laplacian applied and (\ref{DoS4}),
viewed as complex functions of $y^2$.

\section{Ginibre orthogonal ensemble}
\label{s6}
For a reason that will be mentioned below,
we consider the bosonic replicated generating function
(negative moment of the characteristic polynomial)
for the case of Ginibre orthogonal ensemble:
\bea
&&Z_{-n}(z,\z)=
\nn\\
&&\int_{\R^{N \times N}}\!\!\!\!\!\!dH\,\e^{-\frac12 \tr H H^T}
\det^{-n}(z-H) \det^{-n}(\z-H),
\label{Z4}
\eea
$dH=(2\pi)^{-N^2}\prod_{i,j}^N d H^{ij}.$
Again we introduce enlarged flavor indices
$A, B=1,\ldots,2n$ and a $2n\times 2n$ matrix $Z$.
To calculate  
a negative power of the determinant (\ref{Z4})
with the help of  commuting variables $\phb, \phi$,
one has to ensure convergence by
choosing ${\rm Im}\,z>0$ and
inserting a matrix $\sqrt{s}$ defined as \cite{SW} 
\[
\s= \left[
\ba{cc} \openone_n & 0\\
0 & i \openone_{n}
\ea
\right] ,
\]
as a result: 
\widetext
\Lrule
\bea
Z_{-n}(z,\z)
&=&\int_{\R^{N \times N}}\!\!\!\!\!\!dH\,\int_{\C^{2n\times
N}}\!\!\!\!\!\!\! \dbb\,\exp\left(
-\frac12 H^{ij} H^{ij}
+i \phb_{A}^i \s_{AB}
(Z_{BC} \delta^{ij}-\delta_{BC}H^{ij})
\s_{CD}\phi_{D}^j
\right) \nn\\
&=&\int_{\C^{2n\times N}}\!\!\!\!\!\!\! \dbb\,\exp\left(
-\frac12
\phb_{A}^i s_{AB} \phi_{B}^j\phb_{C}^i s_{CD} \phi_{D}^j
+i\phb_{A}^i \s_{AB} Z_{BC} \s_{CD} \phi_{D}^i
\right) \nn\\
&=&\int_{{\cal M}}dQ d\Qt\,
\e^{-\frac12 \tr Q \Qt}
\int_{\C^{2n\times N}}\!\!\!\!\!\!\! \dbb\,
\exp\left(\frac{i}{2}
[\phb_A^i\ \phi_A^i]
\s_{AB}
\left[
\ba{cc}
Q_{BC} & Z_{BC} \\
Z_{BC} & \Qt_{BC}
\ea
\right]
\s_{CD}
\left[
\ba{c}
\phb_D^i\\
\phi_D^i
\ea
\right]
\right) 
\label{58}\\
&=&\int_{{\cal M}}dQ d\Qt\,
\e^{-\frac12 \tr Q \Qt}
{\rm det}^{-\frac{N}{2}}
\left[
\ba{cc}
Q & Z \\
Z & \Qt
\ea
\right]  .
\nn
\eea
\narrowtext
\noindent
Here $d\phb d\phi=(2\pi)^{-2Nn}\prod_{A,i}d^2\phi^i_A$, 
$dQ d\Qt=(2\pi)^{-2n^2-n}\prod_{A\geq B}^{2n} d^2 Q_{AB} d^2 \Qt_{AB}$, 
and the integration domain ${\cal M}$ is a set of pairs of
$2n\times 2n$ complex symmetric matrices $(Q, \Qt)$ that are mutually
Hermitian conjugated, $\Qt=\Qd$.  The restriction to symmetric matrices
is because their antisymmetric parts decouple from the bosonic
bilinears.

Hereafter we concentrate on the center of the circle, 
$|z|\ll \sqrt{N}$.
The large--$N$ saddle point equations
\beq
\frac12 \Qt+\frac{N}{2}\Qt (Q \Qt)^{-1}=0 ,\ \
\frac12 Q+ \frac{N}{2}(Q \Qt)^{-1}Q=0
\eeq
imply 
\beq
Q \Qt=-N\openone_{2n}.
\eeq
Clearly this saddle point manifold lies outside ${\cal M}$, for
$Q \Qt$ would have to be positive definite.
For this saddle point to be admissible,
one must deform
the original integration domain 
${\cal M}=\{Q^+_{AB} \in \R , Q^-_{AB} \in i \R\}$,
$Q^{\pm}\equiv (Q\pm\Qt)/2$, 
into 
$\widetilde{\cal M}=
\{Q^+_{AB} \in \R, Q^-_{AB} \in{\cal C}_{AB} \}$, 
where the contour 
${\cal C}_{AB}$ 
is shown in Figure 1.
\begin{figure}
\epsfxsize=244pt
\begin{center}
\leavevmode
\epsfbox{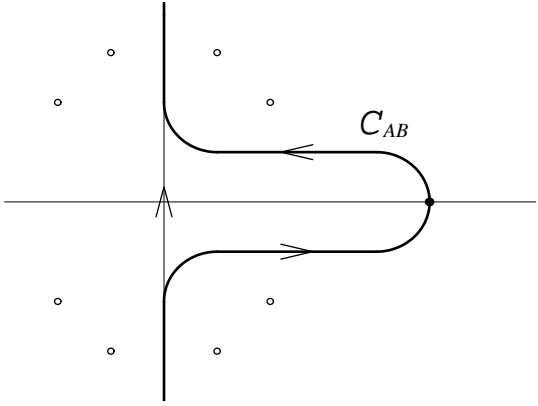}
\end{center}
\caption{
Schematic view of the
integration contours in the complex $Q^-_{AB}$ plane.
The original contour (imaginary axis) can be deformed to
${\cal C}_{AB}$ that passes through the saddle point ($\bullet$)
without encircling the poles ($\circ$) of
the determinant, each of which appears to the left of the
saddle point and symmetrically in four quadrants.
}
\end{figure}
\noindent
The width of the deformed section of the contour scales as 
$1/\sqrt{N}$.
Therefore, as long as $y\gg 1/\sqrt{N} \to 0$, 
the poles of $\det^{-N/2}\left[ {Q\ Z \atop Z\ \Qt} \right]$ 
are away from the deformed contour.
Then the
saddle point equations can be solved by 
\bea 
&&Q=\sqrt{N}U,\ \
\Qt=-\sqrt{N} U^{-1}, \label{spe}\\
&&U\in {\rm GL}(2n,\R), \ \ U=U^T. \nn 
\eea 
For $z$ small and finite of order unity, one can write
for such $Q$ and $\Qt$, 
\bea 
&&\det \left[ \ba{cc} Q & Z \\
Z & \Qt
\ea
\right]
\simeq
{}\exp 
\left(
-\frac1{2N}\,{\rm tr}
\left[
\ba{cc}
0 & \Ui Z\\
-U Z & 0
\ea
\right]^2
\right)\nn\\
&&=
{}\exp \left(\frac1N\left(
2n x^2 -
y^2 \tr U s \Ui s  \right)\right),
\eea
where
$z=x+i y,\ 
Z=x\openone_{2n} +i y s. $

After having deformed the contour of the
integration over $(Q, \Qt)$,
one must re-ensure convergence of the $\phi$
integrals in Eq.\ (\ref{58}) by keeping the real part of the
quadratic form
\bea
&&\frac{i}{2}
[\phb\ \phi] 
\s
\left[
\ba{cc}
Q & Z \\
Z & \Qt
\ea
\right]
\s
\left[
\ba{c}
\phb\\
\phi
\ea
\right]
\nn\\
&=&-y \sum_{A=1}^{2n} |\phi_A|^2 +i x \sum_{a=1}^n(|\phi_a|^2-|\phi_{a+n}|^2)
\nn\\
&&+
[{\rm Re}\,\phi\ {\rm Im}\,\phi]
\s
\left[
\ba{rr}
i Q^+ & Q^- \\
Q^- & -i Q^+
\ea
\right]
\s
\left[
\ba{c}
{\rm Re}\,\phi\\
{\rm Im}\,\phi
\ea
\right] ,
\eea
to be negative definite.
A sufficient condition is
\bea
&&(\s Q^+ \s)_{AB},\  i(\s Q^- \s)_{AB}\in {\bf R}
\nn\\
& &\Leftrightarrow\ 
s Q^\pm s=\pm Q^\pm \ \Leftrightarrow \ 
Q s= s\Qt .
\eea
This condition is actually
necessary for the real part of
the quadratic form to be
negative definite for 
arbitrarily small $y>0$.
At the saddle point $Q=-\Qt^{-1}\equiv \sqrt{N} U$,
it implies a constraint
$U s U s= -\openone.$
Therefore,
\bea
&&Z_{-n}(z,\z)=
\e^{n(N- x^2)}
\int_{{\cal D}}
dU\,
\exp\left(
\frac{y^2}{2}
\tr U s \Ui s
\right),
\label{D}
\\
&&
{\cal D}=\{ U\in {\rm GL}(2n,{\bf R})\, |\, U=U^T,\ (U s)^2 = 
-\openone\},
\label{constraint}
\eea
up to an irrelevant constant factor
that approaches unity in the replica
limit.  Here $dU$ is a Haar measure on the
symmetric space of  symmetric ${\rm GL}(2n,{\bf R})$ matrices,
consisting of $2n+1$ disconnected components
with $2n-k$ positive and $k$ negative eigenvalues.
The group quotient ${\rm GL}(2n,\R)/{\rm O}(2n)$,
which is the saddle point manifold of
the bosonic replicated $\sigma$ model for the
chiral Gaussian orthogonal ensemble,
is isomorphic to the $k=0$ component
by a canonical 
projection $g\mapsto U=gg^T$, but the component used here 
as the integration domain ${\cal D}$
is a subspace of the $k=n$ component.
This difference between non-Hermitian and chiral Hermitian
ensembles is not surprising, because the symmetry breaking
parameter $z$ 
plays a critical r\^{o}le as a regularization parameter,
either in the bosonic replica method or
in the bosonic part of the supersymmetry method in general.

Eq.\ (\ref{D}) is invariant under
\beq 
U\to \left[ \ba{cc} u & 0\\
0 & u'
\ea
\right]
U
\left[
\ba{cc}
u^T & 0\\
0 & {u'}^T
\ea
\right],\ \ u, u'\in {\rm O}(n),
\eeq
so the saddle point manifold shrinks down to the intersection of
${\cal D}$ and ${\rm GL}(2n,\R)/{\rm O}(n)\times {\rm O}(n)$.
A generic solution to the constraint (\ref{constraint}) is
given by
\bea 
U &=&
\left[ \ba{cc} u & 0\\
0 & v
\ea
\right]
\left[
\ba{cc}
\sinh {\bf \Theta}& \cosh {\bf \Theta}   \\
\cosh {\bf \Theta}  & \sinh {\bf \Theta}
\ea
\right]
\left[
\ba{cc}
u^T & 0\\
0 & v^T
\ea
\right] ,
\label{Ureal}\\
&&{\bf \Theta}={\rm diag}(\theta_a),\ u,v \in {\rm O}(n) .
\nn
\eea
Comparing Eq.\ (\ref{Ureal}) with (\ref{Usymm}),
the Haar measure $dU$ in terms of this parametrization is
formally obtained from Eq.\ (\ref{dU}) by
a replacement ${\bf\Theta}\to\pi/2-i{\bf\Theta}$.
Accordingly, 
\beq
dU=du\,dv\,
\prod_{a=1}^n d \cosh\theta_a\, |\Delta(\sinh^2 \theta)|.
\eeq
By substituting
\beq
\tr U s \Ui s=-\tr U^2
= -4\sum_{a=1}^n  \sinh^2\theta_a - 2n\, ,
\eeq
one finally obtains
\beq
Z_{-n}(z,\z)= 
\e^{n(N-|z|^2)}
\int\limits_{0}^\infty\!\! \prod\limits_{a=1}^n
\left( \frac{ d\lambda_a}{ \sqrt{1+\lambda_a} } \,
\e^{-2y^2 \lambda_a} \right)
|\Delta(\lambda)| \, ,
\label{z1}
\eeq
where $\lambda_a=\sinh^2 \theta_a$.
Note that an identical expression also follows from the case with 
$y={\rm Im}\,z<0$.
Once again we have encountered the resemblance between the DoS
(\ref{DoS1}) and the
generating function (\ref{z1}).

For $y^2\gg 1/N$ this integral is dominated by  the vicinity of the
end point $\lambda_a =0$.  By temporarily approximating
$(1+\lambda_a)^{-1/2}\simeq 1$, one obtains 
\beq 
Z_{-n}(z,\z)\simeq {}
\e^{n(N-|z|^2)} t^{-{n/ 2} },\ \ \ \ t=2y^2>0.  
\eeq 
In order to obtain the
perturbative corrections in $1/t$, one expands the
$(1+\lambda_a)^{-1/2}$ factor in powers of $\lambda_a$ as in Sect.\
\ref{s5}.  Accordingly the generating function reads 
\bea
Z_{-n}(z,\z)&=& {} \e^{n(N-|z|^2)}\,
(1-\partial_t)^{-1/2}t^{-{n/ 2}}.  
\eea 
The DoS is then given by 
\bea
\pi R_1(z)&=&\lim_{n\to 0}
\frac{1}{-n} \partial_z\partial_{\z} Z_{-n}(z,\z) \nn\\
&=&
1+\frac12 \{\partial_t ,t\} \partial_t
(1-\partial_t)^{-1/2} \log t .
\label{DoS1rep}
\eea
On the other hand, one can evaluate the asymptotic expansion
of the smooth part of the
exact result (\ref{DoS1}) for ${\rm Re}\,y^2
\gg1$ (this calculation is formally equivalent to the above given one
for $n=1$), 
\bea 
\pi R_1(z)= t(1-\partial_t)^{-1/2}t^{-1}
\ \ \ \ \ \  (t>0).
\label{DoS1ex}
\eea
One can confirm the equality of Eqs.\ (\ref{DoS1rep}) and
(\ref{DoS1ex}) either by using
$[\partial_t, t]=1$ or
by explicitly deriving an identical asymptotic series
\beq
\pi R_1(z)=
\sum_{k=0}^\infty \frac{(-1)^k (2k-1)!!}{(2y)^{2k}}.
\eeq 
The formal Borel resummation of this series leads to the
smooth part of the exact result (\ref{DoS1}).  The $\delta$ function
peak of the DoS localized at $y=0$ cannot directly be captured by our
bosonic replica treatment that requires a finite imaginary part of 
$z$\cite{fn4}.  
However it can be restored from the Borel-reconstructed smooth part
through a normalization sum rule
\beq 
\int\limits_{-\infty}^{\infty}  \bigl( \pi R_1(y) -1 \bigr) dy =0, 
\eeq
with a correct coefficient $\sqrt{\pi}$.
Therefore the bosonic replica
method correctly gives all orders of the perturbative terms
as well as the {\em absence} of
exponentially decaying terms (and the Stokes line at the positive real
$y^2$ axis),
and indirectly the $\delta$ function term of the DoS.

Finally we comment on the fermionic replica treatment of
the orthogonal ensemble.
Following the same steps as in Sect.\ \ref{s5}, one may obtain Eq.\ 
(\ref{UUT}) with the integration domain
being $2n\times 2n$ {\em anti}\,symmetric  unitary matrices,
which is isomorphic to a group quotient ${\rm U}(2n)/{\rm Sp}(n)$ by a
canonical projection $g\mapsto U=g J g^{T}$ (accompanied by a trivial
change $2N\to N$).  At present we are not aware of an appropriate
parameterization, such as Eqs.\ (\ref{Usymm}) or (\ref{Ureal}), of the
intersection of this coset and ${\rm U}(2n)/{\rm U}(n)\times {\rm U}(n)$. 
Consequently these generating functions are not available in the form
of $n$-fold integrals as in Eqs.\ (\ref{z2i}), (\ref{z2ii}),
(\ref{Zn4}), and (\ref{z1}) that are suitable for the asymptotic
analyses.  This difficulty also arises in the bosonic replica
treatment of the symplectic ensemble, whose saddle point manifold is
an intersection of ${\rm U}^*(2n)/{\rm Sp}(n)$ and ${\rm U}^*(2n)/{\rm
U}^*(n)\times {\rm U}^*(n)$.  It is currently not clear to us whether
this difficulty is merely a technical obstacle or has an essential
reason behind it.

\section{Discussions}
\label{s7}
We have derived  the fermionic or bosonic
replicated nonlinear $\sigma$ models for Ginibre
unitary, symplectic, and orthogonal ensembles.
The corresponding symmetric manifolds are
subspaces of ${\rm U}(2n)$, ${\rm U}(2n)/{\rm O}(2n)$, and 
of a space adjacent to ${\rm GL}(2n,\R)/{\rm O}(2n)$ correspondingly, 
in conformity
with Zirnbauer's
classification.  The proper parameterization of the manifolds and
subsequent integration of irrelevant degrees of freedom lefts one with
the $n$--fold compact or noncompact angular integral.  Performing the
asymptotic analysis of such integral, one obtains sums over variety of
possible saddle points both replica-symmetric and replica
nonsymmetric.  After analytical continuation, $n\to 0$, only
replica-symmetric contribution and possibly few nonsymmetric one
survive.  The former are responsible for the perturbative (polynomial
in inverse large parameter), while the latter for the nonperturbative
(essentially singular) contributions to the correlation functions. 
For the DoS calculation, it is possible to evaluate the infinite order
perturbative expansions near these saddle points.  One may also show
that the Borel resummation of such expansions after $n\to 0$ limit
leads to the exact results valid for arbitrary spectral parameter.

We recapitulate our findings for the non-Hermitian ensembles at hand.
The bulk of the DoS in the unitary case (constant) is fully given by 
the
replica-symmetric saddle point. The nonsymmetric contributions show 
up only
if one approaches the very edge of the spectral support, giving the
exponentially 
small deviations from the abrupt behavior on the border of the
circle. The irreducible part of the two-point correlation function
(exponentially
decaying with the distance between the eigenvalues) is fully given by 
a
single 
replica nonsymmetric saddle point.
In the symplectic case the situation is different.
In addition to the symmetric contribution that is determined
to all polynomial orders,
the nonsymmetric contribution exists already on the
level of the bulk DoS. It brings  an exponentially small
correction (not oscillatory as in the case of Dyson ensembles) to the
symmetric contribution,
which is crucial in making contact with the exact result.
In the orthogonal case,
there only exists a single replica-symmetric saddle point that gives 
all
polynomial orders of the DoS,
and the exponential corrections to the symmetric contribution are absent.
One can restore the $\delta$ functional peak from the smooth part of
DoS reconstructed by the Borel resummation.  The presence/absence of
exponential corrections is a direct consequence of
compact/noncompactness of the saddle point manifold which is the
target space of the fermionic/bosonic replicated nonlinear $\sigma$
models.  

Finally we list possible lines of future investigation:\\
\noindent
$\bullet$ The resemblance generally observed
between replicated generating functions
(\ref{z2i}), (\ref{Zn4}), (\ref{z1}) and
corresponding DoS (\ref{DoS2fin}), (\ref{DoS4}), (\ref{DoS1}),
as well as a surprising small--$n$ reduction (\ref{wkw1}) of the 
formers to
essentially one-component integrals, makes us speculate that the
replica treatment may be applicable to Ginibre ensembles beyond
Borel resummation or
asymptotic analyses (for large symmetry breaking parameters 
$|\omega|$, $y$
or at large $N$) 
and could indeed produce full exact results.  Presumably it
could be done by applying a complete set of loop equations to the
$n$-fold integrals to reduce them directly to one-fold
integrals\cite{fn3}.\\
\noindent
$\bullet$
Application to off-diagonal parts of the
remaining two Altland--Zirnbauer superconducting chiral
classes\cite{AZ}, i.e.\ non-Hermitian  
complex symmetric and complex selfdual
 matrices\cite{Has}, should be interesting in its own right, as
analytic results have not yet been obtained via other methods.  It
would require appropriate parameterization of the FF or BB blocks of
the Riemannian symmetric superspaces D$|$C and C$|$D.\\
\noindent
$\bullet$
Starting from  Hatano--Nelson  disordered Hamiltonian,
either with strong or weak non-Hermiticity\cite{FKS,Efe},
one may obtain a  finite-dimensional
nonlinear $\sigma$ model of the corresponding symmetry class.
Our formulation and  results should
serve as a ground upon which non-Hermitian
disordered Hamiltonians in the diffusive regime
(with or without  interactions)  may be treated.

\acknowledgments
We are indebted to A. Altland and M. Zirnbauer for valuable advises 
and conversations.
S.\ M.\ N.\ thanks M. Moshe for warm hospitality at Technion.  This work
was supported in part (S.\ M.\ N.) by the Israel Science Foundation (ISF)
and (A.\ K.) by the BSF-9800338 grant.

\appendix
\section{Jacobian}
\label{a}
In this Appendix
we compute a Jacobian associated with the parameterization
(\ref{Usymm}) following Ref.\cite{ZH}, Sect.\ VIIF and App.\ A.,
where it was proven that

(Zirnbauer-Haldane): \ \ Let
$G$ be a Lie group,
$G_e$ a subgroup that commutes with $s$,
$K$ another subgroup whose right action on $G$ leaves
a function $f$ of $G$ invariant,
$A$ a maximal Abelian subgroup
for the Cartan decomposition of $G$ with respect to $K$,
$M$ a subgroup of $G_e \cap K$ that commutes with all elements of 
$A$, and
$A^+$ an open subset of $A$ such that the map
$\phi: (G_e/M)\times A^+ \to G/K,\ (gM,a)\mapsto gaK$ is bijective.
Let 
$dg$ be a Haar measure on $G$,
$dg_K$ a Haar measure on $G/K$, and
$da$ a Euclidean measure on $A$.
Let $T(\bullet)$ denote a Lie algebra for a Lie group and
a tangent space at the origin for a group quotient.
$T(G)=T(G/K)\oplus T(K)$
decomposes into the even (${}_e$) and odd (${}_o$) elements
with respect to an 
involuntary 
automorphism $g\mapsto sgs$.
As the action of ${\rm ad}(\log a)$, $a\in A^+$, maps
$T(K)_e + T(G/K)_o \to T(G/K)_o + T(K)_e$ and
$T(K)_o + T(G/K)_e \to T(G/K)_e + T(K)_o$,
the set of
positive roots $\Delta^+$ of $G$
decomposes into $\Delta^+_e$ and $\Delta^+_o$ accordingly.
Then the Jacobian $J(a)$ associated with the map $\phi$,
\beq
\int_{G/K}f(gK)dg_K = \int_{A^+}
\left(\int_{G_e} f(g aK)dg\right) J(a)da,
\eeq
is given by
\beq
J(a)=
\prod_{\alpha\in \Delta^+_e} \sinh \alpha(\log a)
\prod_{\alpha\in \Delta^+_o} \cosh \alpha(\log a).
\eeq

In the case of fermionic replica for Ginibre symplectic ensemble,
we take
\bea
&&
G={\rm U}(2n),\ K={\rm O}(2n),\
G_e={\rm U}(n)\times {\rm U}(n),
\nn\\
&&
A \ni
a=\exp \left(
\frac{i}{2}
\left[
\ba{cc}
0 & \bf{\Theta}\\
\bf{\Theta} & 0
\ea
\right] \right)=
\left[
\ba{cc}
\cos({\bf \Theta}/2) & i \sin ({\bf\Theta}/2)\\
i \sin({\bf \Theta}/2) & \cos({\bf\Theta}/2)
\ea
\right],
\nn\\
&&
{\bf{\Theta}}={\rm diag}(\theta_a),\
da=\prod_{a=1}^n d\theta_a ,
 \\
&&f(gK)=\exp\left(-\frac{y^2}{2} \tr gg^T s (gg^T)^{-1} s \right) ,
\label{ggT}
\eea
and identify $gg^T$ in Eq.\ (\ref{ggT})
with $U$ in Eq.\ (\ref{UUT}).
For $g\in G_e$,
\widetext
\renewcommand{\theequation}{A\arabic{equation}}
\bea
&&\tr (ga)(ga)^T s \bigl((ga)(ga)^T\bigr)^{-1} s
=\tr a^2 s a^{-2}  s
\nn\\
&=&
\tr \left[
\ba{cc}
\cos\bf{\Theta} & i \sin \bf{\Theta}\\
i \sin \bf{\Theta} & \cos\bf{\Theta}
\ea
\right]
\left[
\ba{cc}
\openone&0\\
0&-\openone
\ea
\right]
\left[
\ba{cc}
\cos\bf{\Theta} & -i \sin \bf{\Theta}\\
-i \sin \bf{\Theta} & \cos\bf{\Theta}
\ea
\right]
\left[
\ba{cc}
\openone&0\\
0&-\openone
\ea
\right]
=2\sum_{a=1}^n \left( \cos^2 \theta_a-\sin^2 \theta_a \right),
\label{A4}
\eea
and thus the integration over ${G_e}$ is trivial.
Without loss of generality one may restrict $\{\theta\}$
in a cell $\theta_1\geq \theta_2 \geq \cdots\geq\theta_n \geq0$.
The eigensystem of 
${\rm ad}(\log a) X\equiv [\log a, X]=\alpha(\log a)X$ is given by
\bea
\underline{\mbox{Positive Root}}
~~~~~~~~~~~~~~~~~~~~~~~~~~~~~~~~~~~~~\;
&&  
~~
\underline{\mbox{Eigenvector}}
\nn\\
\alpha=i \theta_a\ \ \ \ \ \ \ \in
\Delta^+_o\ \ \ \ \ \ \ \  (a=1,\ldots,n)
\ \ \ \ \,
&&
\ \ 
X_{AB}=\d^{(+)}_{Aa}\d^{(-)}_{Ba}
~~~~~~~~~~~~~~~~~~\,
\in 
T(G/K)_e + T(K)_o\ \ ,
\nn\\
\alpha=i \frac{\theta_a\pm\theta_b}{2}\in
\left\{
\ba{l}
\Delta^+_o\\
\Delta^+_e
\ea
\right.
\ \ \ (1\leq a<b\leq n)
\ \ 
&&
\ \ 
X_{AB}=
\left\{
\ba{ll}
\d^{(+)}_{Aa}\d^{(\mp)}_{Bb}+\d^{(\pm)}_{Ab}\d^{(-)}_{Ba}
& \in 
T(G/K)_e + T(K)_o\\
\d^{(+)}_{Aa}\d^{(\mp)}_{Bb}-\d^{(\pm)}_{Ab}\d^{(-)}_{Ba}
& \in 
T(G/K)_o + T(K)_e
\ea
\right. ,
\eea
where $\d^{(\pm)}_{Aa}\equiv\d_{Aa}\pm\d_{A\,a+n}$.
Accordingly,
\bea
J(a)da&=&
\prod_{a=1}^n d\theta_a
\prod_{a=1}^n \cos \theta_a
\prod_{a<b}^n \left(
\cos\frac{\theta_a-\theta_b}{2}
\cos\frac{\theta_a+\theta_b}{2}
\sin\frac{\theta_a-\theta_b}{2}
\sin\frac{\theta_a+\theta_b}{2}
\right)
\nn\\
&=&
\prod_{a=1}^n d\sin \theta_a
\prod_{a>b}^n \left(
\cos^2 \theta_a-\cos^2 \theta_b
\right) =
\prod_{a=1}^n d\sin \theta_a\,\Delta(\cos^2 \theta),
\eea
\Rrule
\narrowtext
\noindent
up to an irrelevant positive constant.
In a generic cell, one would have
\beq
J(a)da=\prod_{a=1}^n d\sin \theta_a\,|\Delta(\cos^2 \theta)|.
\label{Ja}
\eeq

\renewcommand{\theequation}{B\arabic{equation}}
\renewcommand{\thesection}{B}
\section{Loop equation}
\label{b}
In this Appendix
we derive loop equations\cite{Mig}
for an (auxiliary) Laguerre orthogonal ensemble and solve them in the
replica limit, as it is essential in deriving asymptotic series for 
the
generating functions\cite{DV}.

Consider a set of $n$ random  non-negative numbers $\{\lambda_a\}$
whose unnormalized 
joint probability distribution is given by
\beq 
d\rho(\lambda)=
\prod_{a=1}^n d\l_a \, \e^{-t\l_a} \, |\Delta(\lambda)|.  \eeq Let
$\langle \cdots\rangle$ denote an average with respect to
$d\rho(\lambda)$.
An expectation value of a product of elemantary symmetric polynomials
$\omega_p=\sum_a \l_a^p$,
\[
\bigl\langle \prod_{i=1}^s \omega_{p_i}\bigr\rangle,
\ \ p_i\in {\bf N},
\]
is called a loop amplitude.
Loop equations ({\em a.k.a.}\ Virasoro constraints\cite{FKN}) are
derived from integrals of total derivatives
\beq \int\limits_0^\infty\!\!
\prod_{a=1}^n d\l_a \, \sum_{b=1}^n \frac{\partial}{\partial \l_b}
\left( \l_b^{p_0+1} \prod_{i=1}^s \omega_{p_i} \prod_{c=1}^n
\e^{-t\l_c} |\Delta(\lambda)| \right)=0
\eeq ($p_i=0,1,\ldots$).
By
applying derivatives to each of the factors, one obtains
\bea
&&
\frac{p_0+1}{2}
\bigl\langle \omega_{p_0} \prod_{i=1}^s
\omega_{p_i}\bigr\rangle +\frac12 \sum_{p=0}^{p_0} \bigl\langle
\omega_{p_0-p}\, \omega_p \prod_{i=1}^s \omega_{p_i}\bigr\rangle \nn\\
&&
+\sum_{i=1}^s p_i  
\bigl\langle \omega_{p_0+p_i} \prod_{j(\neq i)}^s  
\omega_{p_j}\bigr\rangle
=t\,\bigl\langle \omega_{p_0+1} \prod_{i=1}^s 
\omega_{p_i}\bigr\rangle .
\label{loop}
\eea
Starting from $\omega_{0}=n$,
this set of loop equations recursively determines
whole loop amplitudes.  First few of them are
\bea &&\langle\o_1
\rangle= {\frac{n ( 1 + n ) }{2 t}}, \nn\\
 &&\langle\o_2 \rangle= {\frac{n ( 1 + 2 n + n^2 ) }{2 t^2}},
 \nn \\ 
 &&\langle\o_1^2 \rangle= {\frac{n ( 2 + 3 n + 2 n^2 + n^3 )
   }{4 t^2}}, \nn\\
 &&\langle\o_3 \rangle= {\frac{n ( 8 + 19 n + 16 n^2 + 5 n^3 )
   }{8 t^3}}, \nn\\
 &&\langle\o_1\o_2 \rangle= {\frac{n ( 4 + 9 n + 7 n^2 + 3 n^3 + n^4 )
   }{4 t^3}}, 
 \nn\\ 
 &&\langle\o_1^3 \rangle=
    {\frac{n ( 8 + 14 n + 13 n^2 + 9 n^3 + 3 n^4 + n^5 ) }{8
    t^3}},\ldots \, .
\eea
Degree-counting in Eq.\ (\ref{loop}) leads to
\beq
\bigl\langle \prod_{i=1}^s \omega_{p_i}\bigr\rangle= t^{-\sum_{i=1}^s
p_i} \left( \upsilon_{p_1,\ldots,p_s} n+ O(n^2) \right)
\label{ups}
\eeq
for $p_i=1,2,\ldots$\, .

Now we prove a surprisingly simple formula
\beq
\upsilon_{p_1,\ldots,p_s} =
\frac12 
\biggl(
\Bigl(\sum_{i=1}^s p_i\Bigr)-1
\biggr)! \, ,
\ \ p_i=1,2,\ldots ,
\label{lemma}
\eeq
by induction:
Eq.\ (\ref{lemma}) holds at degree 1, $\upsilon_1=1/2$.
Assume that it holds for all loop amplitudes of
total degrees $\sum_{i=0}^s p_i = k$.
An amplitude of the degree $k+1$, $\upsilon_{p_0+1,p_1,\ldots,p_s}$
with $p_0+1, p_1, \ldots, p_s\geq 1$ is given as
a linear combination of these amplitudes
by the loop equation (\ref{loop}).
If $p_0\geq1$, its LHS gives
\bea 
\upsilon_{p_0+1,p_1,\ldots,p_s} &=&
\Bigl(\frac{p_0+1}{2}+\frac{p_0-1}{2}+\sum_{i=1}^s p_i \Bigr) \frac12
\Bigl(\sum_{i=0}^s p_i-1\Bigr)!  \nn\\
&=&
\frac12 \Bigl(\sum_{i=0}^s p_i\Bigr)!\, ,
\eea
thus Eq.\ (\ref{lemma}) also holds for
$\upsilon_{p_0+1,p_1,\ldots,p_s}$.  If
$p_0=0$, the first two terms of the LHS of Eq.\ (\ref{loop}) are of
higher orders in $n$, so
one again has 
\beq 
\upsilon_{1,p_1,\ldots,p_s}
\!= \Bigl(\sum_{i=1}^s p_i \Bigr) \frac12 \Bigl(\sum_{i=1}^s
p_i-1\Bigr)!  = \frac12 \Bigl(\sum_{i=1}^s p_i\Bigr)!\, .
\eeq
Therefore Eq.\ (\ref{lemma}) holds for any loop amplitude of the total
degree $k+1$.  End of proof.  It means that the loop amplitude depends
only on the total degree of the symmetric polynomials in the 
small--$n$
(replica) limit\cite{fn2},
\beq
\bigl\langle \prod_{i=1}^s \omega_{p_i}\bigr\rangle=
\bigl\langle \omega_1^{\sum_{i=1}^s p_i}\bigr\rangle
\bigl( 1+O(n) \bigr) .
\label{wkw1}
\eeq
As long as ${\rm Re}\,t>0$ and ${\rm Im}\,t<0$, one can rotate the
integration contours of $\l$'s from $[0,\infty)$ to $[0,i\infty)$
without ever modifying the loop equations and violating convergence.
This justifies the transition from Eq.\ (\ref{wk}) to (\ref{w1}).

\widetext

\begin{references}
\bibitem{fn1}
The upper bound estimated in Eq.\ (15.1.35)
of Ref.\cite{Meh},
which was shown to converge to $\e^{-u^2}/(2\sqrt{\pi}u)$,
is not saturated in
a region of the parameter $1\ll u^2\ll{N}$.
Applying the saddle point approximation
as employed in Sect.\ IIIB, to the integral representation of the
finite-$N$ result (\ref{DoS2fin}),
one obtains the correct asymptotics
(\ref{DoS2e}).
\bibitem{fn4}
In the $y\to 0$ limit, complex conjugate pairs of poles  in
Fig.\ 1
merge at the real axis, so the integration domain ${\cal M}$
cannot be deformed to $\widetilde{\cal M}$ without picking up the
residues.  This is the origin of the 
$\delta(y)$ function peak of the DoS
(\ref{DoS1}).
\bibitem{fn3}
Note that in this paper we have not utilized a loop equation at
$p_0=-1$ (the $L_{-1}$ constraint), which relates
an $n$-fold integral to an $(n-1)$-fold integral arising from the lower
end of the partial integration.
\bibitem{fn2}
This result
is a peculiarity of an orthogonal ensemble
and does not hold for auxiliary ensembles with other powers
of the Vandermonde determinant.
See e.g.\ Ref.\cite{DV}, Eq.\ (94) for a counterexample.
\end{references}
\end{document}